# Distribution Market as a Ramping Aggregator for Grid Flexibility Support

Alireza Majzoobi, Mohsen Mahoor, Amin Khodaei
Dept. of Electrical and Computer Engineering
University of Denver
Denver, CO, USA
Alireza.Majzoobi@du.edu, Mohsen.Mahoor@du.edu, Amin.Khodaei@du.edu

***Abstract*— The growing proliferation of microgrids and distributed energy resources in distribution networks has resulted in the development of Distribution Market Operator (DMO). This new entity will facilitate the management of the distributed resources and their interactions with upstream network and the wholesale market. At the same time, DMOs can tap into the flexibility potential of these distributed resources to address many of the challenges that system operators are facing. This paper investigates this opportunity and develops a distribution market scheduling model based on upstream network ramping flexibility requirements. That is, the distribution network will play the role of a flexibility resource in the system, with a relatively large size and potential, to help bulk system operators to address emerging ramping concerns. Numerical simulations demonstrate the effectiveness of the proposed model on when tested on a distribution system with several microgrids.**

## NOMENCLATURE

*Indices:*

| | |
|---|---|
| $b$ | Index for buses. |
| $f$ | Superscript for fixed loads. |
| $j$ | Index for segments of the load/ramping bids. |
| $m$ | Index for microgrids. |
| $r$ | Superscript for responsive loads. |
| $t$ | Index for time (hour). |

*Parameters:*

| | |
|---|---|
| $c$ | Marginal cost of dispatchable units. |
| $d$ | Load demand |
| $D$ | The demand awarded from the ISO to the DMO. |
| $D^f$ | Total fixed load of all microgrids in the distribution network. |
| $DX^{max}$ | Maximum segment quantity. |
| $\varepsilon$ | Small positive number. |
| $J$ | Number of segments in demand bid. |
| $M$ | Large positive number. |
| $\Delta$ | The desired ramping of the utility grid. |

*Variables:*

| | |
|---|---|
| $DX$ | The amount of load awarded to each bid segment. |
| $PD^M$ | Assigned demand to microgrids by the DMO. |
| $RR$ | Ramping rate of dispatchable units. |
| $RR^{Sel}$ | Selected ramping rate of each microgrid. |
| $RR^{Total}$ | Total ramping capability of all microgrids. |
| $\delta$ | Binary variable representing the selected bid segment. |

## I. INTRODUCTION

BALANCING electricity supply and demand is the most important responsibility of power system operators. The increasing penetration of renewable energy resources which produce a variable generation, however, has challenged the traditional practice in ensuring this balance. Numerous factors, including but not limited to the current renewable portfolio standards, the dropping cost of solar technology, environmental issues, and the state and governmental incentives are boosting this radical change in the U.S. [1]-[3]. The California Independent System Operator (CAISO) has investigated and proposed the "duck curve", in which forecasts that this growing proliferation of renewable energy resources cannot be addressed in near future unless significant levels of ramping are available in the system. The CAISO, and other system operators, are consistently seeking efficient methodologies in order to tackle this emerged issue [4]. To address this problem, power system operators have conventionally used fast response and dispatchable large-scale power generation units. However, these units could potentially become limited by the possible network congestion and are commonly expensive and time-extensive to be built.

In order to investigate the problem of renewable generation integration, two various approaches including large-scale [5]-[8] and small-scale [9]-[12] can be found in the literature. The study in [13] provides a real-time pricing method, by generalizing the concept of ramping costs, to manage a sharp change in electricity demand. By leveraging flexible resources (thermal units, energy storage, and demand response), a stochastic day-ahead scheduling is presented in [14] to manage the variability of renewable generation resources. Moreover, the study in [15] proposes a framework in which the effect of ramping costs on the optimal dispatch of conventional generators is analyzed. The impact of ramping cost in the hourly scheduling of thermal units is studied in [16] via a decomposition-based method.

Reaping the benefits of microgrids with the goal of providing flexibility in the distribution networks is another fruitful strategy to address this issue. Due to significant increase in microgrids deployments in recent years, it is anticipated that, sooner or later, a network of interconnected microgrids will be appearing in power systems [17], [18]. As a result, microgrids can further be utilized for providing flexibility services in distribution network [19]-[23]. The study in [20] leverages microgrids to propose a viable and localized approach to the challenge of distribution network net-load ramping. In the proposed model, microgrid ramping capability is specified through a min-max optimization, then the ramping obstacle is tackled by coordinating both microgrid and distribution grid net loads. Authors in [21] develop a flexibility-oriented microgrid optimal scheduling model to address the ramping issue. This model is expanded on the basis of coordinating microgrid net load with aggregated consumers/prosumers net load, while taking both inter-hour and intra-hour net load variabilities into account. Recently, the variability and uncertainty natures of solar generation are captured by utilizing microgrids [22]. The conclusions drawn from these research efforts advocate that power system operators can considerably take advantage of microgrids to provide flexibility in distribution networks to address the flexibility-associated bottlenecks.

An important issue regarding the capability of microgrids for providing flexibility is to determine the value of ramping in a way that correctly represents microgrid capability and marginal costs. The value of ramping should be determined by the microgrid operator based on a cost-benefit analysis that whether or not it is economical to participate in a distribution market program. An effective valuation approach for microgrid ramping is proposed in [23]. In this respect, two problems, including price-based and ramping-oriented optimal scheduling are defined, and further the microgrid value of ramping is quantified via a comparison between these two problems.

The rapid deployment of microgrids, as well as other proactive customers in distribution networks, has made the case for extending the concept of a Distribution System Operator (DSO) to manage the interaction of these customers with the upstream network as well as with the wholesale market [24]. The existing literature in this research area lacks studies on the microgrids participation on distribution ramping market. Along with the current trend in proposing electricity markets in distribution networks [24], this paper deals with the distribution ramping support under the concept of a Distribution Market Operator (DMO), which is the equivalent of an ISO but in the distribution level.

The reminder of the paper is organized as follows. Section II presents the outline of proposed model, and further develops the problem formulation. Section III represents numerical simulations to demonstrate the effectiveness of the proposed model applied to a test system. Finally, the conclusions drawn from this paper are provided in Section IV.

## II. Model Outline and Problem Formulation

Microgrids have been already proposed as a flexibility resource for increasing the flexibility of the power system and supporting the utility grid to capture the utility ramping, variabilities, and uncertainties [20]-[22]. In line with demand bidding, microgrids can submit their ramping capability to the DMO at each hour. Fig. 1 shows the schematic diagram which demonstrates the interactions of different involved players including the ISO, the DMO and microgrids in the market.

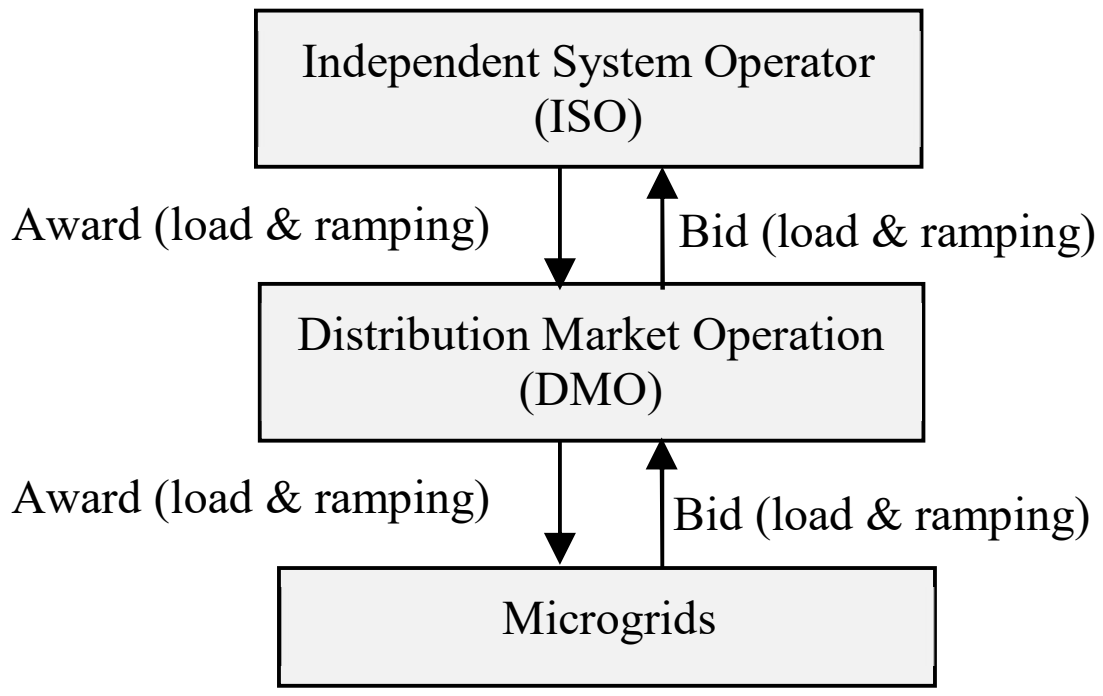

Fig.1. Participation of microgrids in ramping market through the DMO.

As illustrated in the above schematic diagram, the DMO is responsible for two tasks; aggregating microgrids ramping and demand bids and disaggregating the awards from the ISO. In the first step, the DMO combines individual ramping and demand bids received from microgrids, aggregates them, and submits the aggregated bid to the ISO in order to participate in the wholesale energy market. In the second step, the DMO disaggregates the awarded quantity, for both demand and ramping, received from the ISO to microgrids, based on their initially submitted bids. Fig. 2 provides an illustrative example of aggregating ramping curves by the DMO.

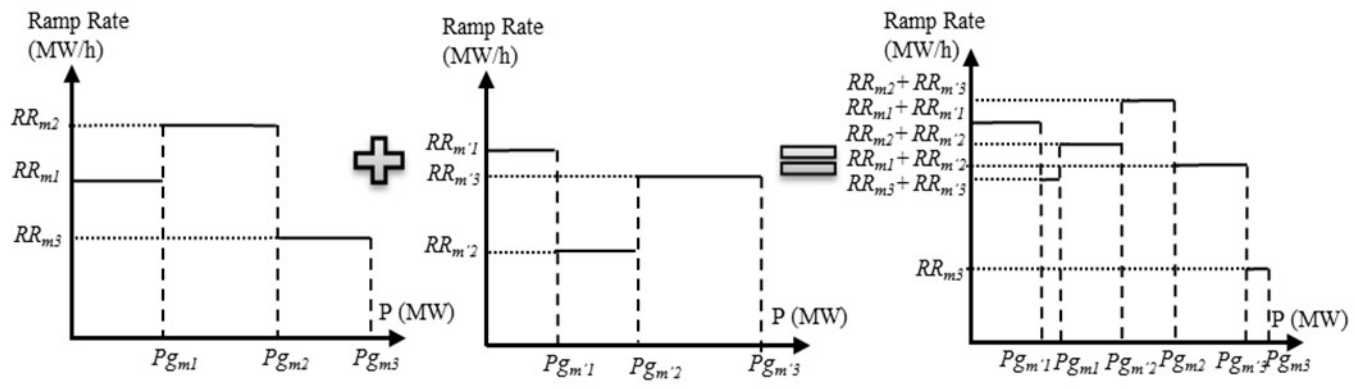

Fig.2. An example of DMO aggregation; two submitted ramping bids by microgrids *m* and *m'* are aggregated in the DMO.

Fig. 3 demonstrates a typical demand bid curve submitted by $m^{th}$ microgrid to the DMO at a sample hour $t$. The fixed part of the loads $d^f$ is not curtailable or shiftable and should be supplied by the utility grid under any circumstances, while the variable part of the bid represents the microgrid flexibility in altering its consumption through load adjustment (which can be done by load curtailment, load shifting, or local generation increase). The summation of all microgrids' fixed loads provides the total fixed load which should be supplied by the DMO (1).

$$D_t^f = \sum_m d_{mt}^f \qquad \forall m, \forall t, \qquad (1)$$

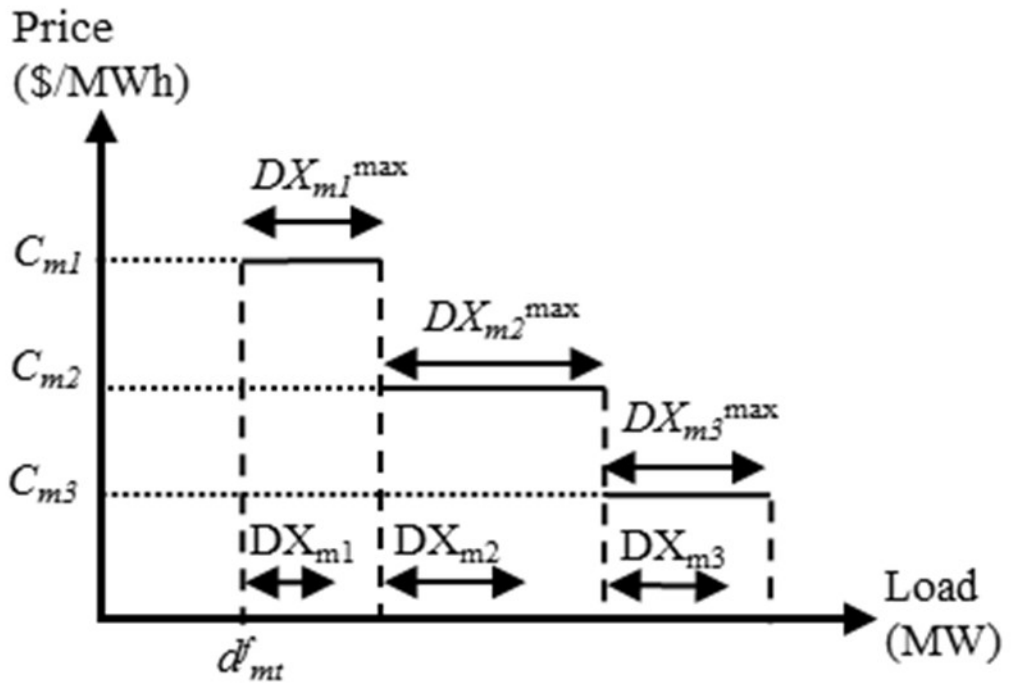

Fig. 3. A typical demand bid curve for microgrid *m*.

After assigning power to the DMO by the ISO, the DMO disaggregates the awarded power to the microgrids. The DMO is aimed at maximizing the demand benefit as in (2) by assigning the optimal awarded power to each microgrid, in accordance with their respective submitted ramping and demand bids.

$$\max \sum_t \sum_m \sum_j c_{mj} DX_{mjt} \tag{2}$$

$$\varepsilon\, \delta_{mjt} \le DX_{mjt} \le DX_{mj}^{\max} \delta_{mjt} \qquad \forall m, \forall t, \forall j, \tag{3}$$

$$\delta_{mjt} \le \delta_{m(j-1)t} \qquad \forall m, \forall t, \forall j, \tag{4}$$

$$d_{mt}^{r} = \sum_j DX_{mjt} \qquad \forall m, \forall t, \tag{5}$$

$$d_{mt}^{r} + d_{mt}^{f} = PD_{mt}^{M} \qquad \forall m, \forall t, \tag{6}$$

$$\sum_m PD_{mt}^{M} = D_{bt} \qquad \forall t. \tag{7}$$

Each load segment is bounded by (3), where binary status variable $\delta$ determines which segments are selected in the optimization model ($\delta_j$ is one when segment *j* has the value of $DX_j$, and it is zero when segment *j* is not selected). Constraint (4) ensures that the segments are selected in a sequential order. The total responsive load of each microgrid equals to the summation of the loads dispatched to each segment (5). The summation of fixed and responsive loads equals to the awarded load to microgrids by the DMO (6). The total demand awarded from the ISO to the DMO is further equal to the summation of the awarded load to microgrids by the DMO (7).

Constraints (8) and (9) are considered in the model to satisfy the desired utility grid ramping at each hour ($\Delta_t$). This desired ramping is supplied by all the participated microgrids in the DSO market. The specific amount of ramping is assigned to each microgrid based on their respective ramping bid.

$$RR_t^{Total} = \sum_m RR_{mt}^{Sel} \qquad \forall t, \tag{8}$$

$$RR_t^{Total} > \Delta_t \qquad \forall t, \tag{9}$$

$$-M(1-\delta_{mjt} + \sum_{k>j} \delta_{mkt}) \le RR_{mt}^{Sel} - RR_{mj} \le$$
$$\le M(1-\delta_{mjt} + \sum_{k>j} \delta_{mkt}) \qquad \forall m, \forall j, \forall t, \tag{10}$$

$$-M \sum_k \delta_{mkt} \le RR_{mj} \le M \sum_k \delta_{mkt} \qquad \forall m, \forall j, \forall t. \tag{11}$$

In order to pair the selected segments of the awarded load with the corresponding ramping capability of those segments, (10) is developed. This constraint is employed for ramping and demand bid curve with J segments, which ensures that by selecting any of the segments of the awarded load, the corresponding ramping value will be selected. On the other hand, if none of the segments are selected, the ramping value will become zero (11).

## III. Numerical Examples

In this section, the proposed model is applied to a test system. A total of 5 microgrids with the total installed DG capacity of 69 MW are considered. Each microgrid consists of 4 dispatchable units with the specifications listed in Table I. Fixed load of each microgrid as well as the total demand awarded to all microgrids from the DMO are plotted in Fig. 4 for 24 hours. It should be noted that as this paper focuses on the role of the DMO and the participated microgrids in the market, a predefined and fixed value is considered as the total demand awarded from the ISO to the DMO.

TABLE I
Marginal Costs ($/MWh), Capacity (MW) and Ramp Rate (MW/h)

| | Price ($/MWh) | | | | |
|---|---|---|---|---|---|
| | MG1 | MG2 | MG3 | MG4 | MG5 |
| DG1 | 71.5 | 62.8 | 64.5 | 69.5 | 76.5 |
| DG2 | 58.4 | 50.5 | 59.8 | 57.2 | 62.4 |
| DG3 | 45.2 | 33.6 | 46.2 | 38.4 | 40.5 |
| DG4 | 23.2 | 25.7 | 27.4 | 27.9 | 31.1 |
| | **Capacity (MW)** | | | | |
| DG1 | 5 | 4 | 5 | 5 | 5 |
| DG2 | 5 | 4 | 3 | 5 | 4 |
| DG3 | 3 | 2 | 3 | 4 | 3 |
| DG4 | 2 | 2 | 1 | 2 | 2 |
| | **Ramping rate (MW/h)** | | | | |
| DG1 | 3 | 2.5 | 3.5 | 2 | 3 |
| DG2 | 2 | 2 | 1.5 | 2 | 1 |
| DG3 | 3 | 2 | 1.5 | 3 | 2 |
| DG4 | 1.5 | 1 | 0.5 | 1 | 1 |

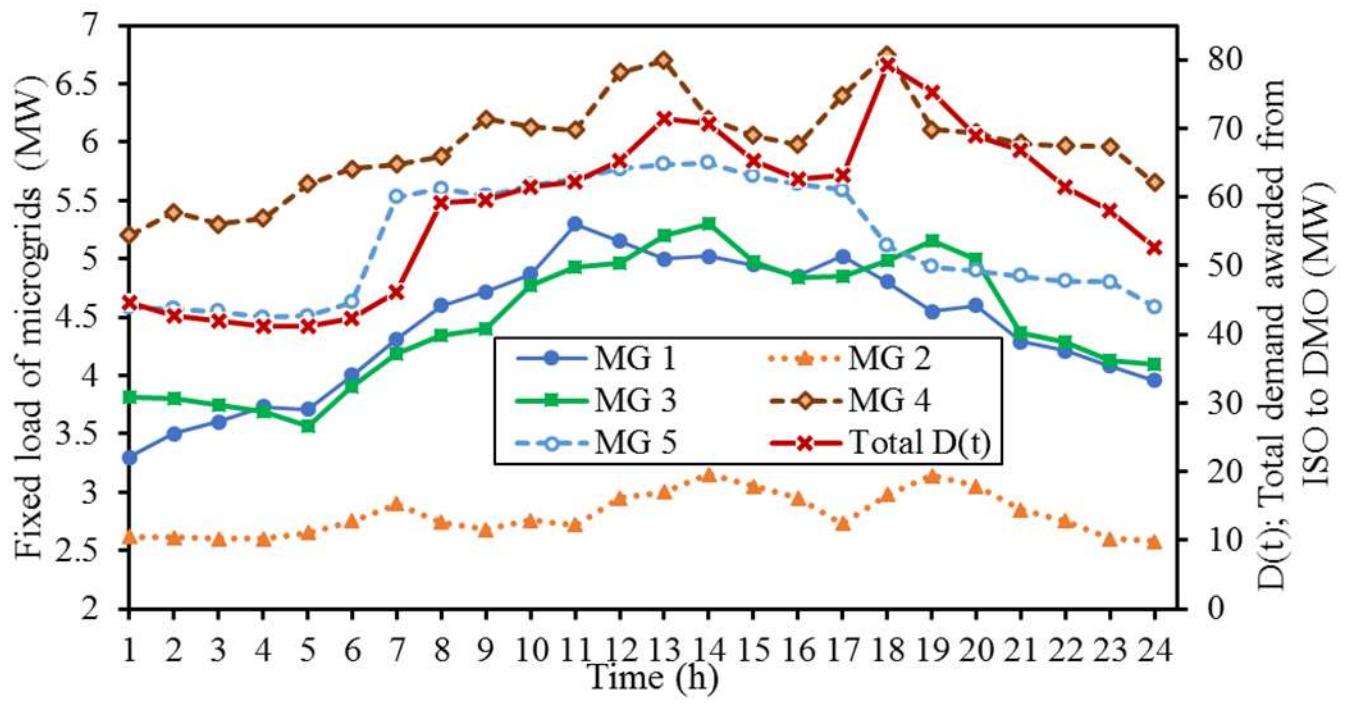

Fig. 4. Fixed load of microgrids and total awarded demand from the ISO to the DMO (MW).

The developed mixed-integer programming problem is solved using CPLEX 12.6. The following cases are studied:

**Case 1**: Market-based microgrid scheduling.

**Case 2**: Market-based microgrid scheduling considering ramping constraints.

**Case 1**: In this case, the load awarded from the ISO to the DMO is distributed between the microgrids based on their bids, while the objective function (2) is maximized. In this case, the DMO does not have any responsibility for providing the ramping to the ISO. The value of awarded load to the microgrids in this case equals $56,286.

**Case 2:** In this case, a total ramping of 12.5 MW/h is considered as the desired ramping value that the DMO is expected to provide to the ISO. The DMO market scheduling problem is solved again with this new constraint. Fig. 5 compares the total ramping capability of all microgrids in Cases 1 and 2. As this figure shows, in Case 2, the participated microgrids provide to at least 12.5 MW/h ramping capability for all hours of the scheduling horizon, which the DMO can reliably deliver to the ISO upon request. This guaranteed ramping should be compared with the available ramping in Case 1, in which is variable, necessarily not guaranteed, and can significantly drop based on microgrids operation.

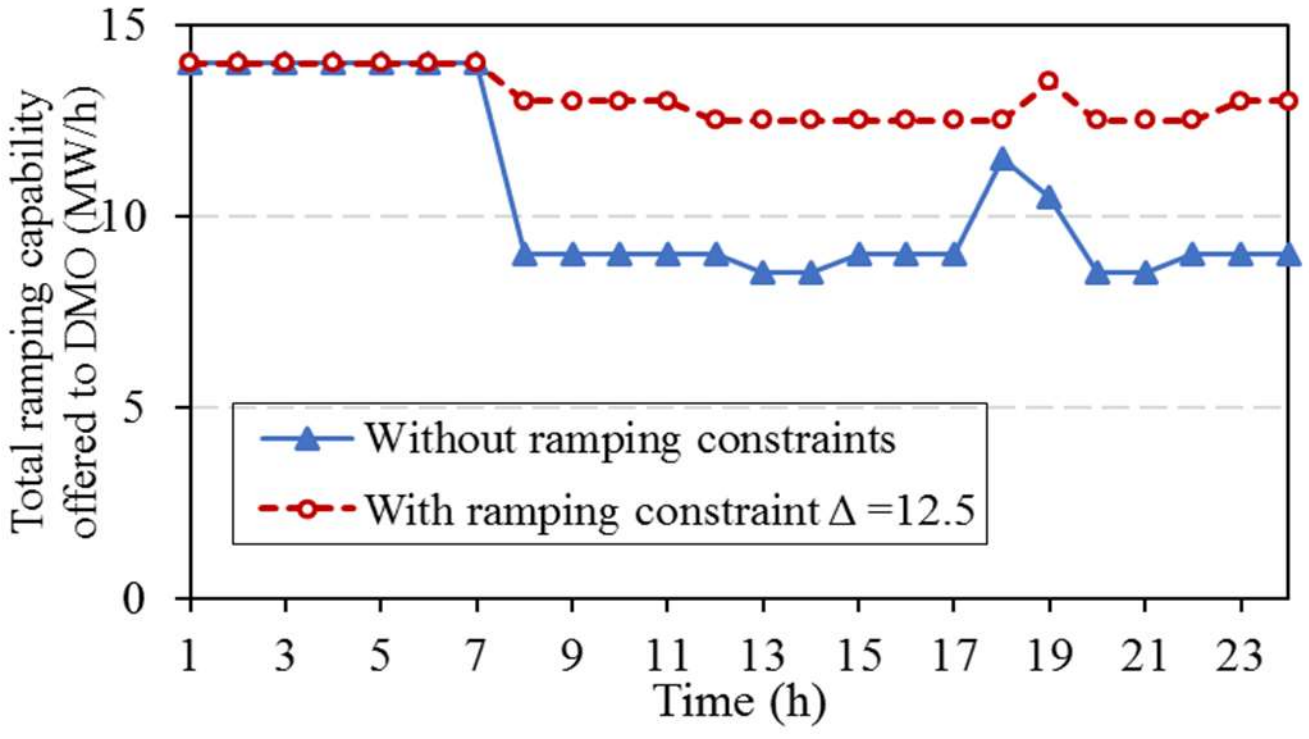

Fig. 5. Total ramping capability of all microgrids offered to the DMO (MW/h).

Fig. 6 compares the awarded load to all 5 microgrids in Cases 1 and 2. Figs 6(a)-6(e) clearly demonstrate how the distribution of the awarded load among the microgrids is alterd based on their ramping and demand bid to the DMO for achieving a 12.5 MW/h ramping capability. For instance, in microgrid 1, Fig 6(a), the awarded load at hours 13, 14, 20, and 21 increases from 10 MW, which leads to moving to the next ramping curve segment with a higher ramping capability (moving from ramp rate of 2 MW/h to 3 MW/h). In Fig. 6 (c) the awarded load to the microgrid 3 decreases in all 24 hours to the first segment (5 MW) which has a larger ramp rate, i.e. 3.5 MW/h. The distribution of the awarded load among the different dispatchable units of microgrids 2, 4 and 5 (Figs 6(b), 6(d), 6(e)) is changed in the same manner to achieve the desired ramping capability. It is intersting to note that the objective value in Case 2 is calcualted as $55,751 ($535 less than Case 1), which is considerably small compared to the significant benefit that the DMO can provide to the ISO.

It is worthwhile to mention that the total awarded load (summation of fixed and adjustable loads) to all microgrids in both cases is exactly the same. However, as Fig. 6 depicts, it is distributed differently among microgrids in order to meet the ramping constraint.

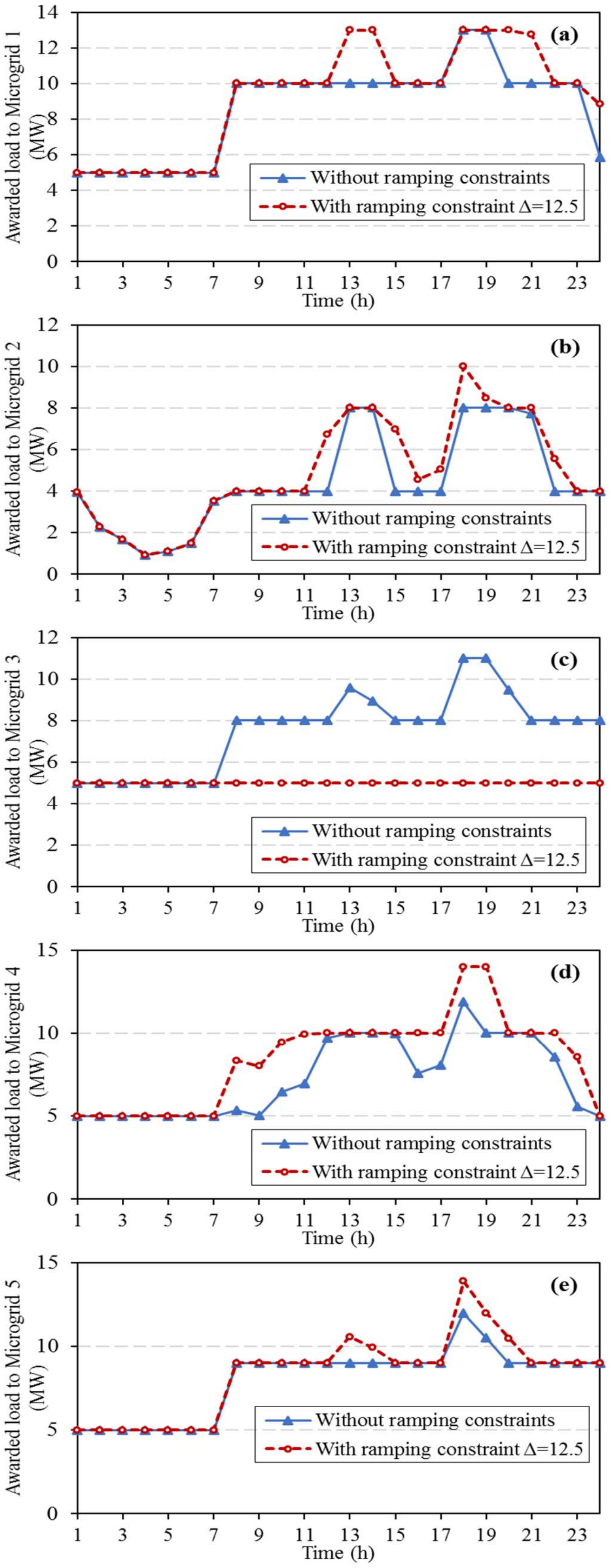

Fig. 6. Comparison of load awarded to all microgrids in two cases with and without ramping constraints.

Table II demonestrates the DG commitments in each microgrid in Case 2, where the highlithed cells represent the changes from Case 1. As the table shows, some of the DG commitments are changed in order to provide the desired ramping to the DMO and consequently to the ISO.

TABLE II
THE UNIT COMMITMENT SCHEDULE OF MICROGRIDS

| | | Time (h) | | | | | | | | | | | | | | | | | | | | | | | |
|---|---|---|---|---|---|---|---|---|---|---|---|---|---|---|---|---|---|---|---|---|---|---|---|---|---|
| | | **1** | **2** | **3** | **4** | **5** | **6** | **7** | **8** | **9** | **10** | **11** | **12** | **13** | **14** | **15** | **16** | **17** | **18** | **19** | **20** | **21** | **22** | **23** | **24** |
| **MG 1** | **DG 1** | 1 | 1 | 1 | 1 | 1 | 1 | 1 | 1 | 1 | 1 | 1 | 1 | 1 | 1 | 1 | 1 | 1 | 1 | 1 | 1 | 1 | 1 | 1 | 1 |
| | **DG 2** | 0 | 0 | 0 | 0 | 0 | 0 | 0 | 1 | 1 | 1 | 1 | 1 | 1 | 1 | 1 | 1 | 1 | 1 | 1 | 1 | 1 | 1 | 1 | 1 |
| | **DG 3** | 0 | 0 | 0 | 0 | 0 | 0 | 0 | 0 | 0 | 0 | 0 | 0 | **1** | **1** | 0 | 0 | 0 | 1 | 1 | **1** | **1** | 0 | 0 | 0 |
| | **DG 4** | 0 | 0 | 0 | 0 | 0 | 0 | 0 | 0 | 0 | 0 | 0 | 0 | 0 | 0 | 0 | 0 | 0 | 0 | 0 | 0 | 0 | 0 | 0 | 0 |
| **MG 2** | **DG 1** | 1 | 1 | 1 | 1 | 1 | 1 | 1 | 1 | 1 | 1 | 1 | 1 | 1 | 1 | 1 | 1 | 1 | 1 | 1 | 1 | 1 | 1 | 1 | 1 |
| | **DG 2** | 0 | 0 | 0 | 0 | 0 | 0 | 0 | 0 | 0 | 0 | 0 | **1** | 1 | 1 | **1** | **1** | **1** | 1 | 1 | 1 | 1 | **1** | 0 | 0 |
| | **DG 3** | 0 | 0 | 0 | 0 | 0 | 0 | 0 | 0 | 0 | 0 | 0 | 0 | 0 | 0 | 0 | 0 | 0 | **1** | **1** | 0 | 0 | 0 | 0 | 0 |
| | **DG 4** | 0 | 0 | 0 | 0 | 0 | 0 | 0 | 0 | 0 | 0 | 0 | 0 | 0 | 0 | 0 | 0 | 0 | 0 | 0 | 0 | 0 | 0 | 0 | 0 |
| **MG 3** | **DG 1** | 1 | 1 | 1 | 1 | 1 | 1 | 1 | 1 | 1 | 1 | 1 | 1 | 1 | 1 | 1 | 1 | 1 | 1 | 1 | 1 | 1 | 1 | 1 | 1 |
| | **DG 2** | 0 | 0 | 0 | 0 | 0 | 0 | 0 | **0** | **0** | **0** | **0** | **0** | **0** | **0** | **0** | **0** | **0** | **0** | **0** | **0** | **0** | **0** | **0** | **0** |
| | **DG 3** | 0 | 0 | 0 | 0 | 0 | 0 | 0 | 0 | 0 | 0 | 0 | 0 | **0** | **0** | 0 | 0 | 0 | **0** | **0** | **0** | 0 | 0 | 0 | 0 |
| | **DG 4** | 0 | 0 | 0 | 0 | 0 | 0 | 0 | 0 | 0 | 0 | 0 | 0 | 0 | 0 | 0 | 0 | 0 | 0 | 0 | 0 | 0 | 0 | 0 | 0 |
| **MG 4** | **DG 1** | 1 | 1 | 1 | 1 | 1 | 1 | 1 | 1 | 1 | 1 | 1 | 1 | 1 | 1 | 1 | 1 | 1 | 1 | 1 | 1 | 1 | 1 | 1 | 1 |
| | **DG 2** | 0 | 0 | 0 | 0 | 0 | 0 | 0 | 1 | **1** | 1 | 1 | 1 | 1 | 1 | 1 | 1 | 1 | 1 | 1 | 1 | 1 | 1 | 1 | 0 |
| | **DG 3** | 0 | 0 | 0 | 0 | 0 | 0 | 0 | 0 | 0 | 0 | 0 | 0 | 0 | 0 | 0 | 0 | 0 | 1 | **1** | 0 | 0 | 0 | 0 | 0 |
| | **DG 4** | 0 | 0 | 0 | 0 | 0 | 0 | 0 | 0 | 0 | 0 | 0 | 0 | 0 | 0 | 0 | 0 | 0 | 0 | 0 | 0 | 0 | 0 | 0 | 0 |
| **MG 5** | **DG 1** | 1 | 1 | 1 | 1 | 1 | 1 | 1 | 1 | 1 | 1 | 1 | 1 | 1 | 1 | 1 | 1 | 1 | 1 | 1 | 1 | 1 | 1 | 1 | 1 |
| | **DG 2** | 0 | 0 | 0 | 0 | 0 | 0 | 0 | 1 | 1 | 1 | 1 | 1 | 1 | 1 | 1 | 1 | 1 | 1 | 1 | 1 | 1 | 1 | 1 | 1 |
| | **DG 3** | 0 | 0 | 0 | 0 | 0 | 0 | 0 | 0 | 0 | 0 | 0 | 0 | **1** | **1** | 0 | 0 | 0 | 1 | 1 | **1** | 0 | 0 | 0 | 0 |
| | **DG 4** | 0 | 0 | 0 | 0 | 0 | 0 | 0 | 0 | 0 | 0 | 0 | 0 | 0 | 0 | 0 | 0 | 0 | **1** | 0 | 0 | 0 | 0 | 0 | 0 |

## IV. CONCLUSIONS

A distribution market scheduling model was proposed in this paper. The proposed scheduling model was developed to capture and collect the ramping capability of participating microgrids in the distribution market as to offer it to the upstream network. Using the proposed model, DMOs can appear as major sources of flexibility in the system to address emerging ramping issues in the system associated with growing proliferation of variable renewable generation. The proposed model was analyzed through numerical simulations, where it was shown that the offered ramping capability could be significant, considering the DMO would collect the ramping capability of a large number of microgrids, and if available, other proactive customers. This offering will be at the expense of minor deviation in microgrids schedules from their optimal operating point, which would require additional discussions on a proper incentive mechanism as follow on work.

## REFERENCES

[1] U.S. Energy Information Administration, "Annual Energy Review," [Online]. Available: https://www.eia.gov/totalenergy/data/annual/previous.cfm.

[2] G. Barbose, N. Darghouth, and S. Weaver, Lawrence Barkley National Laboratory (LBNL), "Tracking the Sun VII: An Historical Summary of the Installed Price of Photovoltaics in the United States from 1998 to 2013," September 2014.

[3] S. Kann, J. Baca, M. Shiao, C. Honeyman, A. Perea, and S. Rumery, "US Solar Market Insight - Q3 2016 - Executive Summary," GTM Research,Wood Mackenzie Business and the Solar Energy Industries Association.

[4] California ISO, "What the duck curve tells us about managing a green grid,". [Online]. Available: https://www.caiso.com/Documents/FlexibleResourcesHelpRenewables_FastFacts.pdf.

[5] A. Ulbig, G. Andersson, "On operational flexibility in power systems," in *Proc. 2012 IEEE Power and Energy Society General Meeting*, pp. 1-8.

[6] E. Lannoye, D. Flynn, and M. O'Malley, "Evaluation of power system flexibility," *IEEE Trans. Power Syst.*, vol. 27, no. 2, pp. 922–931, May 2012.

[7] D. S. Kirschen, A. Rosso, J. Ma, L.F. Ochoa, "Flexibility from the demand side," in *Proc. IEEE Power and Energy Society General Meeting*, San Diego, CA, 22-26 Jul. 2012, pp. 1-6.

[8] M. Ghamkhari and H. Mohmisenian-Rad, "Optimal integration of renewable energy resources in data centers with behind-the-meter renewable generator," in *Proc. IEEE International Conference on Communications (ICC)*, Ottawa, ON, 10-15 Jun. 2012, pp. 3340-3344.

[9] O. Ma et al., "Demand Response for Ancillary Services," *IEEE Trans. Smart Grid*, vol. 4, no. 4, pp. 1988–1995, Dec. 2013.

[10] M. Beaudin, H. Zareipour, A. Schellenberglabe, and W. Rosehart, "Energy storage for mitigating the variability of renewable electricity sources: An updated review," *Energy Sustain. Dev.*, vol. 14, no. 4, pp. 302–314, 2010.

[11] P. Denholm and M. Hand, "Grid flexibility and storage required to achieve very high penetration of variable renewable electricity," *Energy Policy*, vol. 39, no. 3, pp. 1817–1830, 2011.

[12] S. Gottwalt, A. Schuller, C. Flath, H. Schmeck, and C. Weinhardt, "Assessing load flexibility in smart grids: Electric vehicles for renewable energy integration," in *Proc. 2013 IEEE Power and Energy Society General Meeting*, Vancouver, BC, 21-25 Jul. 2013, pp. 1-5.

[13] M. Tanaka, "Real-time pricing with ramping costs: A new approach to managing a steep change in electricity demand," *Energy Policy*, vol. 34, no. 18, pp. 3634-3643, Dec. 2006.

[14] H. Wu, M. Shahidehpour, A. Alabdulwahab and A. Abusorrah, "Thermal generation flexibility with ramping costs and hourly demand response in stochastic security-constrained scheduling of variable energy sources," *IEEE Trans. Power Syst.*, vol. 30, no. 6, pp. 2955–2964, Nov. 2015.

[15] A. J. Lamadrid and T. Mount, "Ancillary services in systems with high penetrations of renewable energy sources, the case of ramping," *Energy Economics*, vol. 34, no. 6, pp. 1959-1971, Nov. 2012.

[16] C. Wang, M. Shahidehpour, "Optimal generation scheduling with ramping costs," *IEEE Trans. Power Syst.*, vol.10, pp. 60–67, Feb. 1995.

[17] Navigant Research, "Microgrid Deployment Tracker 2Q17; Commercial and Industrial, Community, Utility Distribution, Institutional/Campus, Military, Remote, and DC Microgrids: Projects by Region, Segment, and top 10 Countries and Companies," 2017.

[18] S. Parhizi, H. Lotfi, A. Khodaei, and S. Bahramirad, "State of the art in research on microgrids: A review," *IEEE Access*, vol. 3, pp. 890–925, Jul. 2015.

[19] A. Majzoobi and A. Khodaei, "Application of microgrids in providing ancillary services to the utility grid," *Energy,* vol. 123, pp. 555-563, Mar. 2017.

[20] A. Majzoobi and A. Khodaei, "Application of Microgrids in Addressing Distribution Network Net-Load Ramping," *IEEE PES Innovative Smart Grid Technologies Conference (ISGT)*, Minneapolis, MN, 6-9 Sep. 2016.

[21] A. Majzoobi and A. Khodaei, "Application of Microgrids in Supporting Distributed Grid Flexibility," *IEEE Trans. Power Systems*, In press, 2016.

[22] A. Majzoobi, A. Khodaei and S. Bahramirad, "Capturing distribution grid-integrated solar variability and uncertainty using microgrids," in *Proc. 2017 IEEE Power & Energy Society General Meeting,* Chicago, IL, Jul. 2017, pp1-5.

[23] A. Majzoobi, M. Mahoor and A. Khodaei, "Microgrid value of ramping," *IEEE International Conference on Smart Grid Communications*, Dresden, Germany, Oct. 2017.

[24] S. Parhizi, A. Khodaei, M. Shahidehpour, "Market-based vs. price-based Microgrid optimal scheduling", *IEEE Trans. Smart Grid*, In press, 2016.